\documentstyle[prb,aps,twoside,epsf]{revtex}
\begin{document}

\draft

\title{Hall effect in $NS$ and $SNS$ junctions}

\author{F.Zhou and B.Spivak}

\address{Physics Department, University of Washington, Seattle, WA 98195, 
USA}

\maketitle

\begin{abstract}

Hall effect in $SN$ and $SNS$ junctions is considered. It is shown that 
at small temperatures the Hall voltage is 
significantly suppressed as compared to its normal metal value.
The time dependence of the Hall voltage in SNS junctions has a form of 
narrow pulses with the Josephson frequency.

\end{abstract}


Recent progress in microfabrication technology has rekindled the interest
to low temperature proximity effect in superconductor-normal metal ($SN$) 
junctions
 $^{[1-10]}$.
In this paper we consider the Hall effect in superconductor-normal metal 
and superconductor-normal metal-superconductor ($SNS$)junctions.
At low temperatures $T\ll \Delta$, quasiparticles with energies 
$\epsilon\ll \Delta$ can't
 tunnel from the
metal into the superconductor. Here $\Delta$ is the value of the gap in
the superconductors. On the other hand the tunneling of electron pairs,
 which is known as Andreev reflection is possible. This gives
rise to a coherence
between electrons and holes inside the metal, which extends over
distances of order of $L_{\epsilon} =\sqrt{{D}/{\epsilon}} \gg l$.
Here $D=\frac{1}{3}v_{F}l$ is the electron diffusion coefficient, $l$ is the
 elastic mean free
path and $v_F$ is the Fermi velocity. Consequently, the wave packets which
carry the current in the metal
are coherent superpositions of electron and hole wave functions. At $T\ll 
1/\tau={v_F}/{l}$ the packets'
size is of order $L_{T}=\sqrt{{D}/{T}} \gg l$ and their 
effective charge is much smaller than the electron charge $e$. The 
above mentioned electron-hole coherence manifests itself in experimentaly
 observable effects. For example, the density of 
states in the 
metal near the SN boundary $\nu(\epsilon, \vec{r})$ is significantly 
suppressed as compared with 
the bulk metal value $\nu_{0}=m^{2}v_{F}$  $^{[8-10]}$; the low temperature 
resistance of the SN 
junction and the electric field distribution near the SN boundary turn 
out to be very sensitive to the phase breaking rate in the metal $^{[4-7]}$.

 The above mentioned suppression of the effective charge
should also lead to a suppression of the Hall effect. We show below that the 
Hall voltage in SN junction measured with the help of leads 1,2 shown in 
Fig.1a is 
indeed significantly suppressed as compared to its bulk metal value.
We also show that the Hall voltage in SNS junction exhibits oscillations
of its amplitude with Josephson frequency and with no oscillations of the 
sign. It is different fom the behavior of supercurrent through the junction
which at given voltage oscillates both in amplitude and in sign.

In this 
article we neglect the electron-electron interaction in metal and are 
interested only in effects linear in voltage applied to the junction.
In the "clean limit" when $\Delta\tau, T\tau \gg 1$, the kinetic 
scheme describing 
Hall effect in superconductors was developed in $^{[11,12]}$.
In this limit however the above mentioned proximity effect leads only to 
small corrections to the value of the Hall voltage in metal.
To describe the Hall effect in a metal near the SN boundary in the "dirty
limit" when $T \ll \Delta, {1}/{\tau}$ we use the 
following set of equations analogous to that in $^{[13, 
14]}$: $\vec{J}=\vec J_{s}+\vec{J_{n}}+\vec J_{H}$,

\begin{eqnarray}
\vec J_s=\frac{D}{4}
\nu_0 \int d\epsilon f_0(\epsilon) Tr(\sigma_z(G^R\nabla G^R -
G^A\nabla G^A)) \nonumber \\ 
\vec J_{n}=\frac{D}{4}\nu_0\int d\epsilon \nabla f_1(\epsilon) 
Tr(1-\sigma_z G^R\sigma_z 
G^A) \nonumber \\
\vec J_{H}=\frac{1}{16}
\omega_c \tau D\nu_0\int d\epsilon \vec e_b \times \nabla f_1
Tr\{(1+\sigma_z G^R\sigma_z G^R - G^A G^R -\sigma_z G^A\sigma_z G^R)
(G^R\sigma_z-G^A\sigma_z)\},
\end{eqnarray}
\begin{equation}
e\Phi(\vec{r})=
\int d\epsilon f_{1}(\epsilon,\vec{r})\nu(\epsilon,\vec{r})
;\:\:
\nu(\epsilon, \vec{r})=\frac{1}{2}Tr(\sigma_z G^R -G^A \sigma_z),
\end{equation}

\begin{eqnarray}
\epsilon \{G^R\sigma_z - \sigma_z G^R \}+
D \partial \cdot
( G^R\partial G^R)=I_{ph}^R,
\end{eqnarray}

 \begin{eqnarray}
\nabla \cdot \{Tr(1-\sigma_z G^R\sigma_z G^A)\nabla f_1
+Tr(\sigma_z(G^R\partial G^R - G^A\partial G^A))f_0\nonumber \\
+\omega_c\tau Tr((1+\sigma_z G^R\sigma_z G^R - G^AG^R -\sigma_z G^A\sigma_zG^R)
(G^R\sigma_z -\sigma_z G^A))\vec e_b \times \nabla f_1)\}=Tr(\sigma_z
I_{ph}), 
\end{eqnarray}
Here $\vec J_s, \vec J_{n}, \vec J_{H}$ are the supercurrent
density, normal current density and the Hall current density
respectively, $f_0=\tanh(\epsilon/2kT)$ is Fermi distribution function,
$\omega_{c}=eH/mc$ is the cyclotron frequency; $H$ is  
the magnetic field and $\vec e_b$ the unit vector in the direction of magnetic 
field; $\Phi=1/2\partial_t \chi+ \phi$, $\phi$ and $\chi$ 
are the guage invariant scalar potential, electrical potential and 
phase of anomalous Greens function respectively;
$\sigma_{z}$ is the Pauli matrix, $\partial X =\nabla X -i(\sigma_z X 
-X\sigma_z) \vec A$ is the covariant derivative, $\vec A$ is the vector
potential of the magnetic field;
$G^{R,A}$ are retarded and advanced Green 
functions which are matrices in the Nambu space

\begin{equation}
G^{R,A}(\epsilon, \vec r)=
\left (\begin{array}{cc}
g^{R,A} & F^{R,A} \\
- F^{R,A} & g^{R,A}
\end{array} \right ),
\end{equation}
$g^{R,A}(\epsilon, \vec{r})$ and $F^{R,A}(\epsilon, \vec{r})$ are normal and 
anomolous Green functions respectively.
Thus the set of equations consists of electroneutrality condition Eq.2,
the Usadel equation for retarded and advanced Green functions  $G^{R, A}$ 
Eq.3 and the kinetic equation Eq.4 for the distribution function 
$f_{1}(\epsilon,\vec{r})$ which describes the imbalance of 
populations between the electron and hole branches of
spectrum $^{[13,14]}$. 
The scattering integrals $I_{ph}^{R}\approx i\sigma_z\tau_{in}$ and 
$Tr(\sigma_z I_{ph})$ describe the broadening of the spectrum 
and the imbalance charge relaxation
due to inelastic processes. 
In Eqs.3,4, we take into account that in noninteracting metal
the order parameter and the supercurrent are zero.

In the zeroth order approximation in the parameter $\omega_{c}\tau$, these 
equations were derived in $^{[13]}$. However, to describe the Hall
effect one
has to keep terms linear in $\omega_{c}\tau$. The problem of 
the Hall effect in the mixed state of superconductors was addressed 
recently in $^{[12, 15, 18]}$. 
 We derive Eqs.1,3 which include the contributions linear in 
$\omega_{c}\tau$ using the same procedure as in 
$^{[11,12,15, 18]}$.

Let us now consider Hall effect in SN junction shown in Fig.1a.
We assume that the magnetic field is smaller than the critical one and does 
not penetrate into the superconductor. 
Taking into account the  normalization condition
 $(g^R(\epsilon, \vec r))^2= F_2^{R}(\epsilon, \vec r)
F_1^R(\epsilon, \vec r) +1$ one can express $g^R$, $F^R$ and Eqs.1-4 in the 
form

\begin{equation}
\cos\theta(\epsilon, \vec r)=g^R(\epsilon, \vec r), \:\: 
\sin\theta(\epsilon, \vec r) =i F_{1}^{R}(\epsilon, \vec r)
\end{equation}       
\begin{equation}
\frac{D}{2}{\nabla}^2\theta(\epsilon)+
(i\epsilon-\frac{1}{\tau_{in}})
\sin\theta(\epsilon)
-\frac{D}{4}(\frac{e H}{\hbar c})^2 x^2 \sin 2\theta=0
\end{equation}
\begin{eqnarray}
\nabla \cdot \{ \cosh^2\theta_2 \nabla f_1 +\tau \omega_c \vec e_b \times
\nabla f_1 \cos\theta_1 \cosh^3\theta_2 \}=0
\nonumber \\
\vec J=D\nu_0 \int d\epsilon \{\cosh^2\theta_2 \nabla f_1 +\tau\omega_c
\vec e_b\times \nabla f_1 \cos\theta_1 \cosh^3\theta_2\}
\end{eqnarray}
where $\theta=\theta_1 +i\theta_2$ is the
complex variable. We have chosen $\vec A=\vec e_y Hx$. 

The boundary conditions for Eqs.7,8 have the forms$^{[16]}$

\begin{eqnarray}
D\vec n\cdot \nabla\theta(\epsilon, \vec R)=
t\sin(\theta(\epsilon,0^+)-\theta(\epsilon, 0^-))
\nonumber \\
D\cosh\theta^2_2 \cdot\nabla f_1
=t v_F\{f_1(0^{-}) -f_1(0^+)\}\cosh\theta_2(0^+)\cosh\theta_2(0^{-})
\cos(\theta_1(0^+)-\theta_1(0^-))\nonumber \\
f_1(0)=0, f_1(\epsilon, x=\infty)= -eV \partial_\epsilon f_0(\epsilon)
\end{eqnarray}
Here $\vec{n}$ and $t$ are the unit vector perpendicular to the boundary 
and the transmission coefficient of the boundary, respectively. Indecies 
$+, -$ indicate two sides of the $SN$ or $NN_p$ interface (Fig.1). 
Inside the superconductor $\theta(\epsilon \ll \Delta)=\pi/2$ while 
inside the bulk metals $\theta=0$.

 Solving Eqs.8,9 in zeroth 
order approximation in $\omega_{c}\tau$
we get the expressions for the conductance $G_{SN}=\frac{I_0}{U_{0}}$ of 
the junction shown in Fig.1a $^{[4-7]}$ measured by the two probe method and 
the conductance $G_p$ measured between the leads 1,2. (Here $I_{0}$ 
is the current through the junction and
$U_0$ is the voltage across the junction). That is
\begin{eqnarray}
G_{SN}(T)=S_0\sigma_0 \int d\epsilon \partial_{\epsilon}f_0  
\frac{1}{L_S(\epsilon)},
G_p(T)=S_p\sigma_p \int d\epsilon \partial_{\epsilon}f_0
\frac{1}{2L_p(\epsilon)}
\end{eqnarray}
where
\begin{eqnarray}
L_S(\epsilon)=\frac{L_t}{\cosh\theta_2(\epsilon, 0)
\sin\theta_1(\epsilon,0)}+\int^{L_0}_{0} dx \frac{1}
{\cosh\theta_2^2(\epsilon, x)},\nonumber \\
L_p(\epsilon)=\frac{L_{tp}}{\cosh\theta_2(\epsilon, L)
\cosh\theta_{2p}(\epsilon,0)
\cos(\theta_1(\epsilon, L)-\theta_{1p}(\epsilon,0))}+
\int^{L_p}_{0} dy \frac{1}
{\cosh\theta_{2p}^2(\epsilon, y)},
\end{eqnarray}
$S_{0,p}$ are the cross sections of the sample and the leads,  
$\sigma_{0,p}, D_{0,p}$ are 
the Drude conductivities and diffusion constants in the sample and leads;
$L_t =D_0/t_0 v_F, L_{tp}=D_p/t_p v_F$ are the length characterising
$SN$ and $NN_p$ boundaries,
$t_0$ and $t_p$ are the transmission coefficients for $SN$
and $NN_p$ boundaries, $\theta_p(\epsilon,y)$ is the solution of Eq.7 inside 
the leads.

In the first order approximation in $\omega_{c}\tau$ we get an expression 
relating the Hall current $I_H$ and Hall voltage $U_H$ measured by leads 1,2 

\begin{eqnarray}
I_H =G_{p}(T)U_{H}(T)=
\tau\omega_c I_0 \frac{d}{G_{SN}(T)}
\int d\epsilon \frac{S_p\sigma_p}{L_p(\epsilon)}
\frac{1}{L_S(\epsilon)}\frac{\cos\theta_1(\epsilon, L)}
{\cosh\theta_2(\epsilon, L)}\partial_{\epsilon} f_0
\end{eqnarray} 
Here $d$ is the width of the junction (See Fig.1).
The values of $\theta_1, \theta_2$ and $U_{H}$ 
(as well as $G_{SN}$ and $G_{p}$) depend on  
the processes breaking the electron-hole coherence inside 
the sample and in the leads. Therefore, they are very sensitive to the 
ratio between parameters $L_0, L_T, L_{t}, L_{tp}$ 
$^{[4,5,7]}$.

The solution of Eq.7 in the metal for $SN$ junctions when $L_0 \gg L_t $
with the corresponding boundary conditions in Eq.9 is $^{[7]}$ 

\begin{equation}
\theta(\epsilon, L)=\left \{ \begin{array}{cc}
\frac{\pi}{2} -\sqrt{2}\frac{L_t+L}{L_0}  &
\mbox{$L_t, L \ll L_0 \ll L_\epsilon$} \\
\frac{\pi}{2} -(1-i)\frac{L_t+L}{L_\epsilon}  &
\mbox{$L_t,L \ll L_\epsilon \ll L_0$}\\
\frac{\sqrt{2}\pi(1+i)L_\epsilon}{4L_t}
exp(-(1-i)\frac{x}{\sqrt{2}L_\epsilon})&
\mbox{$L_\epsilon \ll L_t \ll L_0$}.
\end{array}\right.
\end{equation} 

The most interesting results appear in the cases when the processes 
which break the electron-hole coherence are not effective and the 
value of $\theta_1(T, L)$ is 
close to $\pi/2$. At small $T$ and large $t$ 

\begin{equation}
\frac{U_{H}}{U_{HN}}=\left \{\begin{array}{cc}
\max\{\frac{L_t}{L_0}, \frac{L_t}{L_T}\}&
\mbox{$L \ll L_t \ll L_T$} \\
\max\{\frac{L}{L_0}, \frac{L}{L_T}\}&
\mbox{$L_t \ll L  \ll L_T$}. 
\end{array}\right.
\end{equation}
Here $U_{HN}=\omega_{c}\tau d{I_{0}}/S_0\sigma_0$ is the Hall 
voltage measured by the leads $1,2$ in the absence of the 
proximity effect.
 The main feature of Eq.14 is that $U_{H}$ is significantly 
suppressed as compared with $U_{HN}$(See Fig.1a).

At high temperatures
when $L_T \ll L_t$ or $L$ the proximity effect gives only small
 corrections to $U_{HN}$
\begin{equation} 
\frac{U_{H} -U_{HN}}{U_{HN}}=
min\{\frac{L_T^2}{L_t^2},
\frac{L^2_T}{L^2}\} \ll 1
\end{equation}

The above results are obtained in the limit of low magnetic field
when $L_{H} \ll L_T$. Here $L_{H}=\sqrt{\Phi_0/H}$ is the magnetic length,
$\Phi_0$ is flux quantum. In the opposite 
limit one should substitute $L_{T}$ for $L_H$ in Eqs.14, 15.
As we have mentioned the value of $U_H$ is very sensitive to the nature of 
the leads.
The requirement that the leads do not contribute to the breaking of the 
electron-hole coherence is ${d} \ll {L_{tp}}, {L_t}\ll {L_0}$.

Let us now turn to the discussion of the Hall effect in $SNS$  
junctions. In this case  the ac Josephson effect causes the values of 
$\theta(\vec{r},\chi_0)$ and 
$\nu(\vec{r},\chi_0)$ to be time dependent. At small $U_0$ 
these 
dependences adiabatically follow the corresponding time
dependence of the order parameter phase difference $\chi_0$ across the
junction: $\partial_t\chi_{0}(t)=2eU_{0}$. 
For simplicity let us consider the case of a thin junction when $L_{1} \ll 
L_H$ when one can neglect any
$y$-dependence of $\chi_{0}$ along the junction. 
It is well known that at $\chi_{0}=0$, the energy dependence of 
$\nu(\epsilon)=\cos\theta_1(\epsilon)\cosh\theta_2(\epsilon)$
 exhibits a gap $E_{g}(\chi_{0}=0)\sim {D}/{L_0^2}$ $^{[3]}$.
The $\chi_0$ dependences of $\theta$, $\nu$ and $E_{g}$ were calculated in 
$^{[10,17]}$.
It was shown that the value of the gap $E_{g}$ monotonically decreases with 
$\chi_0$ and closes at $\chi_{0}=\pi$.  
Hence $\cos\theta_1 (\epsilon)=0$ when
$\epsilon \leq E_g(\chi_0)$.   
In two limiting cases when $\chi_0$ is close to $\pi$ or $0$ 
$E_g(\chi_0)$ is determined by$^{[17]}$ 
 \begin{equation}
E_g(\chi_0)\approx E_{g}(0)\left\{ \begin{array}{cc}
(1-C\chi^2_0) & \mbox{$\chi_0 \ll \pi$} \\
C_1(\pi -\chi_0) & \mbox{$\pi-\chi_0 \ll \pi$}
\end{array} \right.
\end{equation}
where $C$ and $C_1$ are of order of unity.
It follows from Eq.12 that the Hall voltage $U_H$ is a periodical 
function of time with a period $1/2eU_0$. 
At low temperatures $T \ll E_c$ the value of
$U_{H}(t)$ is exponentially small except when the gap is small 
$E_{g}(\chi_0(t))\sim T$. As a result $U_{H}(t)$
consists of short pulses of duration $\tau^* \sim {T}/E_c {eU_{0}}$ 
with maximums of order of $U_{HN}$ (See Fig.1b). Let $t_{n}$ be the time 
when $nth$ maximum of $U_{H}(t)$ occurs. Then at $\tau^* \ll |t-t_n| \ll 
1/2eU_0$, 
\begin{equation} 
U_{H}(t) \sim  U_{HN}\exp(-\frac{|t-t_n|}{\tau^{*}})
\end{equation}
The main contribution to 
the Hall current averaged over the period of oscillations 
comes from the time intervals
$|t-t_n| \sim \tau^{*} \ll 1/U_{0}$ and as a result
\begin{equation}
<U_{H}>\sim \frac{T}{E_c} U_{HN}
\end{equation}
We would like to mention the difference in the temperature dependences in 
$SN$ 
and $SNS$ cases. In first case $U_{H}\sim \sqrt{T}$ while in the second case
$U_{H}\sim T$. 
Let us note that the time dependence of the resistance of the junction 
exhibit peaks which are similar to the above considered peacks of the 
Hall voltage.

In addition to the quasiparticle's contribution to the Hall
current considered above,
there is another contribution which can be associated with the 
supercurrent $^{[15,18]}$. We believe, however, that this part 
of the Hall current does not contribute to the Hall voltage.

In this paper we considered the limit $eU_{0} \ll 
{1}/{\tau_{\epsilon}}$ when it is possible to neglect nonequilibrium 
corrections to the quasiparticle distribution function inside the normal 
metal. In the opposite limit the quasiparticle distribution in the metal 
at $\epsilon <E_c$ becomes nonequilibrium and as a result the duration of 
the pulses $\tau^*$ becomes of order of $(eU_{0})^{-1}$. 

The above results were obtained for the 
quantities averaged over realizations of random potential.
We are planing to consider mesoscopic contributions in $U_H$ elsewhere.

We acknowledge useful discussions with B.Altshuler, A.I.Larkin and 
B.Pannetier. This work was 
supported by Division of Material Sciences, U.S.National Science Foundation
under Contract No.DMR-9625370.

\newpage
\begin{figure}
\begin{center}
\leavevmode
\epsfbox{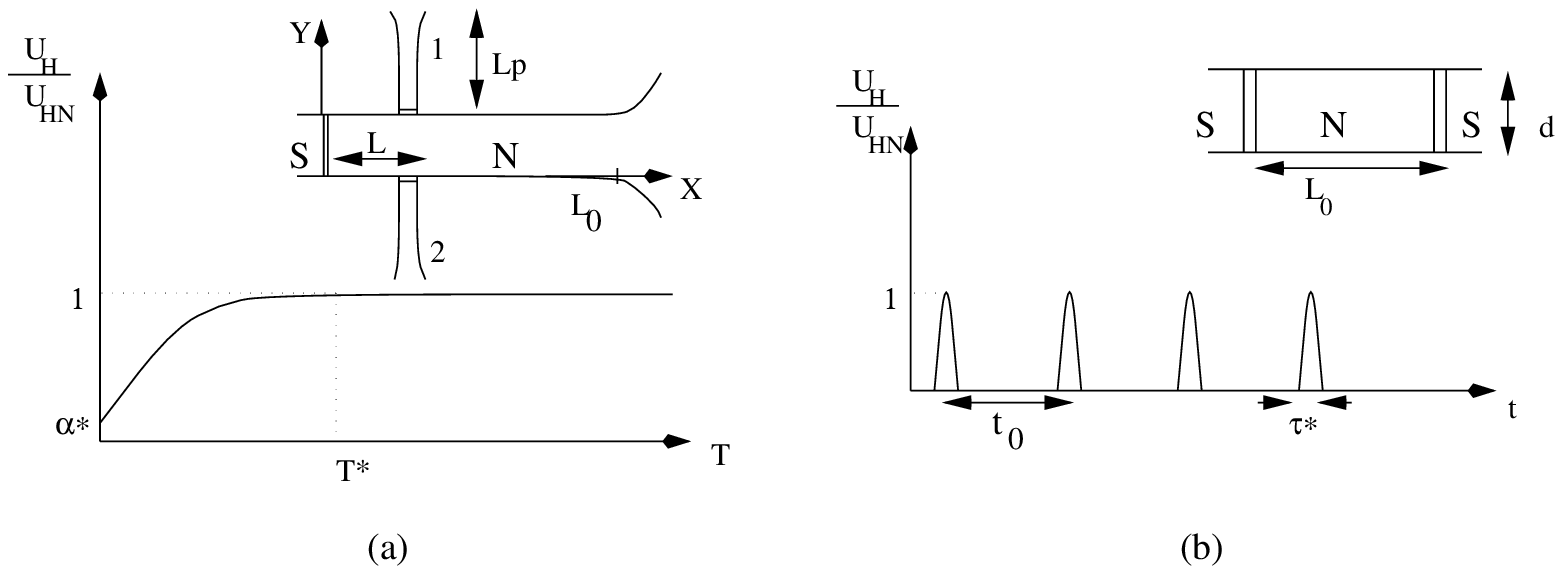}
\end{center}
Fig.1. a). The temperature dependence of Hall voltage in $SN$ junction. 
$\alpha^*=max \{\frac{L_t}{L_0},\frac{L}{L_0}\}$, 
$T^* =min\{\frac{D_0}{L^2}, \frac{D_0}{L_t^2} \}$.
 b). Time dependence of Hall voltage in $SNS$
junctions with geometry shown in the insert, $\tau^*=t_0 T/E_c  
\ll t_0=1/2eU_0$. 
\end{figure}
\end{document}